# The nature of localization in graphene under quantum Hall conditions

J. Martin[1,2], N. Akerman[2], G. Ulbricht[3], T. Lohmann[3], K. von Klitzing[3], J. H. Smet[3] and A. Yacoby[1,2]

[1] *Department of Physics, Harvard University, Cambridge, MA 02138, USA.*

[2] *Department of Condensed Matter Physics, Weizmann Institute of Science, Rehovot 76100, Israel*

[3] *Max-Planck-Institut für Festkörperforschung, Heisenbergstrasse 1, D-70569 Stuttgart, Germany.*

**Particle localization is an essential ingredient in quantum Hall physics [[1],[2]]. In conventional high mobility two-dimensional electron systems Coulomb interactions were shown to compete with disorder and to play a central role in particle localization [[3]]. Here we address the nature of localization in graphene where the carrier mobility, quantifying the disorder, is two to four orders of magnitude smaller [[4],[5],[6],[7],[8],[9],[10]]. We image the electronic density of states and the localized state spectrum of a graphene flake in the quantum Hall regime with a scanning single electron transistor [[11]]. Our microscopic approach provides direct insight into the nature of localization. Surprisingly, despite strong disorder, our findings indicate that localization in graphene is not dominated by single particle physics, but rather by a competition between the underlying disorder potential and the repulsive Coulomb interaction responsible for screening.**

The quantum Hall effect is manifested in transport through vanishing longitudinal resistance and a quantized transverse resistance [1]. In graphene, the relativistic nature of the underlying particles and the absence of valley mixing give rise to quantum Hall phases at filling factors $\nu = 4i - 2$ where the 4 fold degeneracy is a result of both spin and valley degeneracies [4,5,12]. To date it is well established that disorder plays a crucial role in the formation and robustness of quantum Hall phenomena [1,2]. While at low magnetic fields the relativistic nature of the electronic spectrum of graphene prevents particle localization, the application of a strong magnetic field perpendicular to the layer opens up Landau gaps in the spectrum that lead to strong localization [13]. Clearly the universal nature of the quantum Hall effect in macroscopic specimens relies on its insensitivity to the details of the underlying disorder mechanism. Already single particle physics of two dimensional electrons in a disordered potential landscape leads to localization of charge and scaling behavior [1,2]. However, in high mobility GaAs specimens localization is no longer solely driven by disorder and a fundamentally different picture of localization emerges that is strongly influenced by the presence of Coulomb interactions leading to Coulomb blockade and non-linear screening of the disorder potential [3,14,15,16].

Thus far Coulomb interactions in clean two dimensional electron systems are known to have a profound effect and may bring about for example the fractional quantum Hall effect [17] and quantum Hall ferromagnetism. The latter was recently observed in graphene [18,19,20,21,22]. However, the carrier mobility in non-suspended graphene, a common measure of the disorder strength, is typically $10^3$ to $10^4$ times smaller than the mobility characterizing GaAs based 2D systems [4,6,6,7,8]. Even in suspended layers peak mobilities are up to a hundred times smaller than those measured in high mobility GaAs layers [9,10]. It raises the question whether Coulomb interactions play a significant role in particle localization in graphene or perhaps single particle physics suffices. To address this issue, we explore the electronic density of states in graphene at high magnetic fields using a scanning single electron transistor. We focus our work on the quantum Hall regime where we image the bands of localized states both in position and energy. Our findings clearly indicate that at large magnetic fields charge localization is governed by the presence of strong Coulomb repulsion between electrons despite the high level of disorder and the behavior is similar to the one observed in GaAs two dimensional electron systems with their much larger carrier mobilities.

The preparation of graphene monolayers was performed in a similar manner to that reported in Ref. [23,24]. A detailed description is deferred to the method section. The sample was mounted in the preparation chamber of an ultra high vacuum scanning system and subsequently annealed by heating it above 100 °C at a background pressure of $5 \times 10^{-7}$ mbar. The entire annealing procedure removed most of the doping adsorbates introduced during fabrication and exposure to humidity and the neutrality point shifted close to zero back-gate voltage [25]. A back-gate voltage difference of 1V corresponds to a density change of $7 \times 10^{10}$ cm$^{-2}$. This conversion factor was extracted from two-terminal magneto-transport data.

We begin our investigations by measuring the energy level diagram as a function of the applied external magnetic field and density. Using a scanning probe microscope equipped with a single electron transistor (SET) on a glass fiber tip the local chemical potential $\mu$ and the inverse compressibility $\partial \mu / \partial n$ are recorded. The SET measures the local electrostatic potential. However, under equilibrium conditions, any changes in the local chemical potential are exactly compensated for by changes in the local electrostatic potential. Hence, the SET enables to detect the local chemical potential as well as its local derivative $\partial \mu / \partial n$ when weakly modulating the carrier density via the doped Si backgate. The latter quantity is inversely proportional to the local compressibility and directly related to the local density of states. The details of this technique have been described previously in Refs. [11,26]. Figure 1a depicts a schematic of the experimental arrangement with the SET hovering above the monolayer.

The energy spectrum of graphene can be determined by measuring the chemical potential as a function of the carrier density. In a fixed magnetic field and in the absence of disorder, the chemical potential undergoes sudden jumps equal to the energy between adjacent Landau levels. They arise when gradually increasing the density whenever the degeneracy of the highest filled Landau level is exhausted and the next Landau level starts to be filled. The chemical potential can either be measured directly with the SET or is obtained by integrating compressibility data. A color rendition of the inverse compressibility as a function of magnetic field $B$ and carrier density $n$ for a fixed position on the graphene flake is displayed in Fig. 1b. The gaps in the energy spectrum give rise to regions where the inverse compressibility is large. These regions appear as lines at filling factors ν= 4$i$-2 with $i$ = -1, 0, 1, 2.....Fig. 1c depicts a measurement of the chemical potential (top curve) together with a cross section through the inverse compressibility data of Fig. 1b at a fixed magnetic field of 11.7 T. Prominent integer fillings have been marked. For the sake of comparison, the theoretical behavior of the chemical potential and the inverse compressibility at zero magnetic field have been included as well (dotted lines). Substantial disorder broadening [27] in graphene is responsible for the absence of abrupt jumps in the chemical potential. Only a smooth undulation is observed when an additional Landau level gets filled. From these data the Landau level energy spectrum can be reconstructed. A maximum in the compressibility (or a minimum in the depicted inverse compressibility) heralds the center of the individual Landau levels. The chemical potential values at the densities where such compressibility maxima appear yield the Landau level positions on the energy axis. A slight refinement of this simple procedure to extract the energy spectrum is required due to stray electric fields and is discussed in the method section. The resulting Landau level spectrum after the stray field correction is plotted in Fig. 1d. A similar spectrum has been recently obtained from scanning tunneling spectroscopy [28]. The energy of the center of each Landau level agrees well with the expected theoretical value as given by $E_i = sgn(i) \cdot \sqrt{2e\hbar v_F^2 |i| B}$. Here, $i$ is the Landau level index ($i > 0$ for electron like and $i < 0$ for hole-like Landau levels) and $v_F$ the Fermi velocity. A best fit to the data yields a Fermi velocity equal to $1.1 \pm 0.1 \cdot 10^6$ m/s which differs only by 10% from the value expected from band structure calculations [29]. This value is also consistent with our previous compressibility studies in the absence of a magnetic field from which a nearly identical value of the Fermi velocity was obtained [30]. Theoretical work addressing disorder and interactions confirm such weak renormalization of the Fermi velocity [13,31,32].

The color rendition in Fig. 1b provides an overall view of the behavior of the local compressibility. We now zoom into the individual incompressible regions by recording data on a much finer grid of the density versus magnetic field plane. The outcome is depicted in Fig. 2a. Previous works on massive 2D electron systems made out of GaAs have shown that such local compressibility maps are dominated by sharp peaks which run parallel to the lines of constant integer filling when the system can still reach equilibrium on the time scale of the measurement [3]. Similar behavior is seen here in graphene at large magnetic fields. Many dark lines or spikes which run parallel to the line of constant filling factor 2 and -2 are visible. We will argue that these lines represent the charging spectrum of the localized states. Below we show that the number of charging lines and their parallel character constitutes a direct proof that despite the strong disorder, localization in graphene is not a single particle effect but rather involves an elaborate interplay between the disorder potential and local charge reorganization driven by the repulsive Coulomb interaction.

It is instructive to consider the expected behavior of the localized state spectrum for strong disorder where one can neglect the effects of the Coulomb repulsion and screening first. Under these conditions each localized state consists of a constant energy contour of the disorder landscape enclosing an *integer* number of flux quanta. When the magnetic field increases, these constant energy orbits must shrink in size in order to maintain a constant flux through their area and their energy changes accordingly. Fig. 3a displays a random disorder potential and two sets of localized states at fields $B_2 > B_1$. Fig. 3b plots a vertical cross-section through the disorder potential and illustrates how the constant energy orbits shrink as the field is ramped up. In addition the number of localized states should increase at any given location across the sample in proportion to the Landau level degeneracy per unit area since this degeneracy has increased in view of the larger number of flux quanta which now thread the sample. For a particular localized state the density at which it becomes populated depends on the specific details of the disorder landscape. In particular this density evolves not necessarily monotonic in magnetic field. These key features of the single particle picture of localization in the strong disorder regime are illustrated in a cartoon like fashion in Fig. 3c-d. For simplicity the Landau level spin and valley degeneracy's are ignored and the filling of only one Landau level is shown. Panel 3c plots at a given field the average number of occupied states per unit area $n$ in the Landau level when the chemical potential $\mu = E$. Also the density at which the localized states shown in panels a and b get filled are marked.

Finally, panel d highlights the evolution of the two selected localized states in the ($n,B$)-plane. Localized states may cross and they do not run parallel to the line corresponding to complete filling of the Landau level. We conclude that the generic features of single particle localization such as the increase in the number of localized states with increasing field and their erratic, non-monotonic behavior in the ($n,B$)-plane are entirely absent in the data of Fig. 2a. Clearly, this picture does not hold for localization in graphene. We note that while the SET is a local probe it will nonetheless pick up the signature of all localized states within a spatial extent given by its size (of order 100 nm).

The behavior of the localized state spectrum described above should be contrasted with the expected behavior if Coulomb repulsion is strong enough to compete with disorder and to enforce local charge rearrangements to screen the bare disorder potential. Figure 4 illustrates this scenario. Far away from complete filling of the Landau level, the electrons manage to flatten the potential by setting up a spatially varying density profile. They succeed in screening the bare disorder potential because many states are available to rearrange electrons in the partially filled level. However as one approaches an integer filling, the degeneracy $n_{max}$ of the partially filled Landau level may get exhausted and electrons can no longer rearrange to flatten the disorder potential. Antidots or dots, surrounded by these incompressible regions where screening fails, form as illustrated in the left panel of Fig. 4 and Coulomb blockade physics becomes relevant. Dark lines or spikes in the local compressibility would reflect the charging with a single electron of a discrete localized state in one of the dots or antidots in close proximity to the SET. Since the bare disorder remains fixed independent of field, the same landscape of dots and antidots recurs at any field. In the middle panel for instance the magnetic field was increased. The level degeneracy $n_{max}$ has gone up accordingly. The same anti-dots and localized states appear at higher density. Hence, the localized state spectra of the dots/antidots produce parallel lines in the ($n,B$) plane (right panel), rather than erratic, non-monotonic lines as in the single particle model. This behavior of the localized states can serve as a fingerprint of the interacting model for particle localization. It is well pronounced in the data recorded on graphene in Fig. 2a. Moreover, the same set of localized states would also reappear at different filling factors as well. Therefore the number of localized states and the width measured along the density ordinate of the incompressible band where such localized states occur in the Landau fan diagram is expected to remain fixed independent of field and filling factor, provided that Landau levels are well separated. In Fig. 5 we plot the width in density of the band of localized states centered around integer filling where graphene condenses in a quantum Hall state as a function of magnetic field. For a given magnetic field, this width is extracted by fitting a sum of Lorentzian (Gaussian) functions to the inverse compressibility trace as illustrated in Fig. 5a. A Lorentzian (Gaussian) is centered at each integer filling where a maximum appears. The width or variance is assumed identical for all Lorentzians (Guassians). This fitting procedure works well as long as Landau levels are well separated. Fig. 5b clearly demonstrates that as the field increases and the separation between the Landau levels starts to exceed the broadening, the width of the incompressible band is saturating to a constant, magnetic field independent value. It corroborates further that Coulomb interactions and non-linear screening are crucial ingredients for localization in graphene under quantum Hall conditions.

Up to this point, measurements were restricted to a single location. Fig. 2b shows a color rendition of the inverse compressibility recorded along a line across the graphene flake as a function of the average sample density (or equivalently filling factor) at $B=10\,\text{T}$. It reveals the spatial dependence of the charging spectrum and the bottom border of the incompressible band centered around filling factor 2. The charging of each localized state is now seen to persist over a finite spatial extent which is set either by the localization length or by the resolution limit given by the size of the tip (~100nm). One can clearly see that electrons tend to localize at particular locations in space. We anticipate a strong correlation between these local density variations and the border of the incompressible band. It was shown previously that with a scanning SET it is also possible to extract the local density from a differential measurement of the surface potential in the compressible and incompressible regime [30]. The outcome of such a measurement has been included in Fig. 2b as the blue trace. For the incompressible behavior we used Landau filling factor 2 at an average density of about 3.5 $10^{11}/\text{cm}^2$. The surface potential measurement in the compressible regime was performed at the center of the Landau level at the charge neutrality point for a field of 10 T. The correspondence between the density fluctuations and the border of the incompressible band, i.e. the appearance of localized states is striking.

The typical length scale of the density fluctuations in the line scan of fig. 2b is around 150 nm. It is resolution limited by the size of the SET sensor. The intrinsic disorder length scale of graphene was reported previously from the density interval spanned by the incompressible band around any of the well-resolved integer filling factors [30]. It was estimated to be approximately equal to 30 nm. A detailed study of spatial maps of the localized states spectra as in Fig. 2b offers however an alternative approach to estimate the characteristic length scale of the disorder. The level spacing between the localized states from a single dot or antidot is intimately connected with its size. Fig. 2c shows one example of a spectrum of equidistant compressibility spikes that apparently stems from a single dot. It exhibits large level spacing. In order to add one extra electron to this dot the average density in the sample needs to be changed by approximately 3 $10^{10}/cm^2$. Hence the electron is localized within an area with a diameter of approximately 60 nm. This size fits reasonably well with the previously reported estimated length scale from the width of the incompressible band.

**Acknowledgements:**

We would like to acknowledge useful discussions with Yigal Meir and Assa Auerbach. This work is partly supported by the Harvard NSEC.

**Methods**

*1. Sample preparation*

An adhesive tape was used to peel a large graphite flake from an HOPG crystal. It was pressed onto a Si/SiO$_2$ wafer. The conducting Si substrate served as a back-gate to vary the carrier density in the graphene sheet. Once a suitable monolayer was identified in an optical microscope, two electrical contacts were patterned with optical lithography. Typical dimensions of the monolayers were $10 \times 4$ μm². After lift-off of the ohmic contact metal (3 nm Cr, 30 nm Au), the graphene flakes exhibited their neutrality point at back gate voltages larger than 100V. Therefore, the sample surface was cleaned during an Ozone treatment and with an ammonia dip. As-prepared flakes had their neutrality point at back gate voltages near 30 V. The sample was annealed by heating it above 100 °C at a background pressure of $5 \times 10^{-7}$ mbar in the ultra-high vacuum scanning system. This procedure apparently removed the doping adsorbents introduced during fabrication and exposure to humidity as the charge neutrality point shifted close to zero back-gate voltage.

*2. Extraction of the Landau level spectrum*

The procedure to extract the Landau level spectrum requires some refinement as some electrical field lines originating from the back-gate reach the SET. The parasitic charge associated with these fringing fields is also detected by the SET. The detected amount of charge depends both on the distance of the SET to the graphene edge as well as on the vertical distance between the tip and the flake. For the heights and distances to the sample edge used in these experiments, we derive a contribution to the compressibility from the fringing fields of approximately 0.15 meV $10^{-10}$ cm². This direct pickup will add a linear term to the chemical potential as we sweep the backgate voltage. The linear contribution from fringing fields to the chemical potential was determined from fitting the chemical potential data at zero magnetic field to the theoretically expected behavior of the chemical potential with an added linear term. The data in the presence of a magnetic field was then corrected with the linear term stemming from the fringing field.

**Figures:**

Figure 1: The Landau level spectrum in monolayer graphene extracted from compressibility measurements. **a.** Schematic of the device geometry consisting of a graphene monolayer and a source and drain contact on top of a doped Si substrate covered by a 300 nm thick thermal oxide. A scanning SET is hovering above the graphene sample at a distance of a few tens of nm. **b.** Color rendition of the inverse compressibility $\partial \mu / \partial n$ in units of $10^{-10}$ meV·cm² as a function of the applied magnetic field $B$ and the carrier density $n$. **c.** Measured chemical potential $\mu$ (top curves) and inverse compressibility $\partial \mu / \partial n$ at a fixed field $B = 11.7$ T (solid lines) in comparison with the calculated behavior of $\mu$ and $\partial \mu / \partial n$ at zero field (dotted lines). **d.** Landau level energies as a function of the magnetic field extracted from the $\mu$ and $\partial \mu / \partial n$ measurements. The solid lines plot the Landau level energies assuming a Fermi velocity $v_F = 1.1 \cdot 10^6$ m/s.

Figure 2: Measured spectrum of localized states in graphene. **a.** Color rendition of the inverse compressibility measured as a function of average carrier density (controlled using the back gate) and external magnetic field. The measurement is taken at a single location above the flake. Localized states in this measurement appear as dark lines that run parallel to the filling factor. **b.** Color rendition of the inverse compressibility measured along a line across the graphene flake and as a function of average carrier density. Localized states in this measurement appear as dark horizontal lines spanning a small spatial extent determined by the localization length or by the resolution limit of the tip. The solid line indicates the carrier density profile along the same line extracted from the surface potential measurements. **c.** Localized state spectrum of a single dot. The level spacing is inversely proportional to the size of the dot and yields a measured dot size of 60 nm.

Figure 3: Single particle picture of localization in a perpendicular magnetic. For the sake of simplicity, only one non-degenerate Landau level is considered. **a.** Example of a bare disorder potential with hills and valleys. Localized states correspond to orbits of constant energy which enclose an integer number of flux quanta. When increasing the field ($B_2 > B_1$), these constant energy orbits shrink in size in order to maintain constant flux. Their energy may drop or increase depending on the details of the disorder potential. Their number increases in proportion to the increase in the level degeneracy per unit area $n_{max} = eB/h$. **b.** Cross-section through the disorder potential. The localized states shown in **a** are also highlighted here. **c.** Plot of the position of the chemical potential as a function of the average carrier density $n$ in the Landau level for two values of the field $B_1$ and $B_2$. At large values of the chemical potential, the density approaches the level degeneracy $n_{max}$. **d.** Diagram plotting when each localized state becomes populated in the $(n, B)$-plane. Population of a state

causes a spike in the local compressibility.

Figure 4: The behavior of localized states in the presence of non-linear screening due to Coulomb repulsion **a.** Electrons rearrange and set-up a spatially varying density distribution (dotted line) to flatten the bare disorder potential. This is possible as long as the required local charge density is smaller than the level degeneracy: $n_{max}$. Close to complete filling of the level, the level degeneracy may get locally exhausted. The density profile will exhibit isolated pockets of unfilled states or so-called anti-dots as shown here (solid blue line). Their charging involves Coulomb blockade physics and gives rise to compressibility spikes. **b.** At different magnetic fields, the same pockets of unfilled states will recur for the same distance in density from the $n_{max}$ line. **c**. Due to this recurrence, each localized state in the ($B$,$n$)-plane gives rise to a line which runs parallel to the line of complete filling $n = n_{max}$ (or $\nu = 1$).

Figure 5: The magnetic field dependence of the width in density of the incompressible region. **a.** The width of the incompressible region is determined from a best fit to the data. The fit function is composed of Lorentzians centered at the maxima of the inverse compressibility. The variance of the Lorentzian serves as the fit parameter and is taken identical for all Lorentzians. A fit example is shown for the compressibility data recorded at 11.3 T. **b.** The dependence of the density width of the incompressible region on the applied magnetic field. These widths are extracted as described under **a.** A sum of Lorentzians (blue squares) as well as Guassians (red circles) have been used. The solid line corresponds to 4eB/h and demarcates the transition from the regime of separated to overlapping levels.

1  Prange, R. E. & Girvin, S. M. eds. *The Quantum Hall Effect* (Springer, New York, 1987).
2  Huckestein, B. Scaling theory of the integer quantum Hall effect. Rev. Mod. Phys. **67**, 357-396 (1995).
3  Ilani, S., Martin, J., Teitelbaum, E., Smet, J. H., Mahalu, D., Umansky, V., Yacoby, A. The macroscopic nature of localization in the quantum Hall effect. Nature **427**, 328-332 (2004).
4 Novoselov, K. S., Geim, A. K., Morozov, S. V., Jiang, D., Katsnelson, M. I., Grigorieva, I. V., Dubonos, S. V., Firsov, A. A. Two-dimensional gas of massless Dirac fermions in graphene. Nature **438**, 197-200 (2005).
5  Zhang, Y., Tan, Y.-W., Stormer, H. L., Kim, P. Experimental observation of the quantum Hall effect and Berry's phase in graphene. Nature **438**, 201-204 (2005).
6  Geim, A. K., Novoselov, K. S. The rise of graphene, Nature Materials **6**, 183- 191 (2007).
7  Tan, Y.-W., Zhang, Y., Bolotin, K., Zhao, Y., Adam, S., Hwang, E. H., Das Sarma, S., Stormer, H. L., Kim, P. Measurement of scattering rate and minimum conductivity in graphene. Phys. Rev. Lett. **99**, 246803 (2007).
8  Chen, J.-H., Jang, S., Adam, S., Fuhrer, M. S., Williams, E. D., Ishigami, M. Charged impurity scattering in graphene. Nature Physics **4**, 377-381 (2008).
9  Du, X., Skacho, I., Barker, A., Andrei, E. Y. Approaching ballistic transport in suspended graphene. Nature Nanotechnology **3**, 491-495 (2008).
10  Bolotin, K. I., Sikes, K. J., Jiang, Z., Klima, M., Fudenberg, G., Hone, J., Kim, P., Stormer, H. L. Ultrahigh mobility in suspended graphene, Solid State Communications **146**, 351-355 (2008).
11  Yoo, M. J., Fulton, T. A., Hess, H. F., Willett, R. L., Dunkleberger, L. N., Chichester, R. J., Pfeiffer, L. N., West, K. W. Scanning single-electron transistor microscopy: Imaging individual charges. Science **276**, 579–582 (1997).
12  Ostrovsky, P. M., Gornyi, I. V., Mirlin, A. D. Theory of anomalous quantum Hall effects in graphene, Phys. Rev. B **77**, 195430 (2008).
13  Castro Neto, A. H., Guinea, F., Peres, N. M. R., Novoselov, K. S., Geim, A. K. The electronic properties of graphene. Reviews of Modern Physics, accepted for publication.
14  Fogler, M. M., Novikov, D. S., Shklovskii B. I. Phys. Rev. B **76**, 233402 (2007).
15  Fogler, M. M. Neutrality point of graphene with coplanar charged impurities. arXiv:0810.1755 (2008).
16  Struck, A., Kramer, B. Electron correlations and single particle physics in the integer quantum Hall effect, Phys. Rev. Lett. **97**, 106801 (2006).
17  Jain, J. K. Composite Fermions. Cambridge University Press, Cambridge UK (2007).
18  Nomura, K., MacDonald, A. H. Quantum Hall ferromagnetism in graphene. Phys. Rev. Lett. **96**, 256602 (2006).
19  Alicea, J., Fisher, M. P. Interplay between lattice-scale physics and the quantum Hall effect in graphene. Solid State Communications **143**, 504-509 (2007).

20  Checkelsky, J. G., Li, L., Ong, N. P. Divergent resistance at the Dirac point in graphene: evidence for a transition in high magnetic field. arXiv:0808.0906 (2008).

21  Sheng, L., Sheng, D. N., Haldane, F. D. M., Balents, L. Odd-integer quantum Hall effect in graphene: interaction and disorder effects. Phys. Rev. Lett. **99**, 196802 (2007).

22  Pereira, A. L. C., Schulz, P. A. Valley polarization effects on localization in graphene Landau levels. Phys. Rev. B **77**, 075416 (2008).

23  Novoselov, K. S., Geim, A. K., Morozov, S. V., Jiang, D., Zhang, Y., Dubonos, S. V., Grigorieva, I. V., Firsov, A. A. Electric field effect in atomically thin carbon films. Science **306**, 666–669 (2004).

24  Novoselov, K. S., Jiang, D., Schedin, F., Booth, T. J., Khotkevich, V. V., Morozov, S. V., Geim, A. K. Two-dimensional atomic crystals. Proc. Natl. Acad. Sci. USA **102**, 10451–10453 (2005).

25  Schedin, F., Geim, A. K., Morozov, S. V., Hill, E. W., Plake, P., Katselnson, M. I., Novoselov, K. S. Detection of individual gas molecules adsorbed on graphene. Nature Materials **6**, 652–655 (2007).

26  Yacoby, A., Hess, H. F., Fulton, T. A., Pfeiffer, L. N., West, K. W. Electrical imaging of the quantum Hall state. *Solid State Commun.* **111**, 1–13 (1999).

27  Zhu, W., Shi, Q. W., Wang, X. R., Chen, J., Hou, J. G. the shape of disorder broadened Landau subbands in graphene. arXiv: 0810.4475v1 (2008).

28  Li, G., Luican, A., Andrei, E. Y. arXiv:0803.4016 (2008).

29  Brandt, N. B., Chudinov, S. M. & Ponomarev, Y. G. Semimetals 1, Graphite and its compounds (North-Holland, Amsterdam, 1988).

30  Martin, J., Akerman, N., Ulbricht, G., Lohmann, T., Smet, J. H., von Klitzing, K., Yacoby, A. Observation of electron-hole puddles in graphene using a scanning single electron transistor. Nature Physics **4**, 144-148 (2008).

31  Barlas, Y., Pereg-Barnea, T., Polini, M., Asgari, R., MacDonald, A. H. Chirality and correlations in graphene. *Phys. Rev. Lett.* **98**, 236601–236604 (2007).

32  Hu, B. Y.-K., Hwang, E. H., Das Sarma, S. Density of states of disordered graphene. Phys. Rev. B **78**, 165411 (2008).

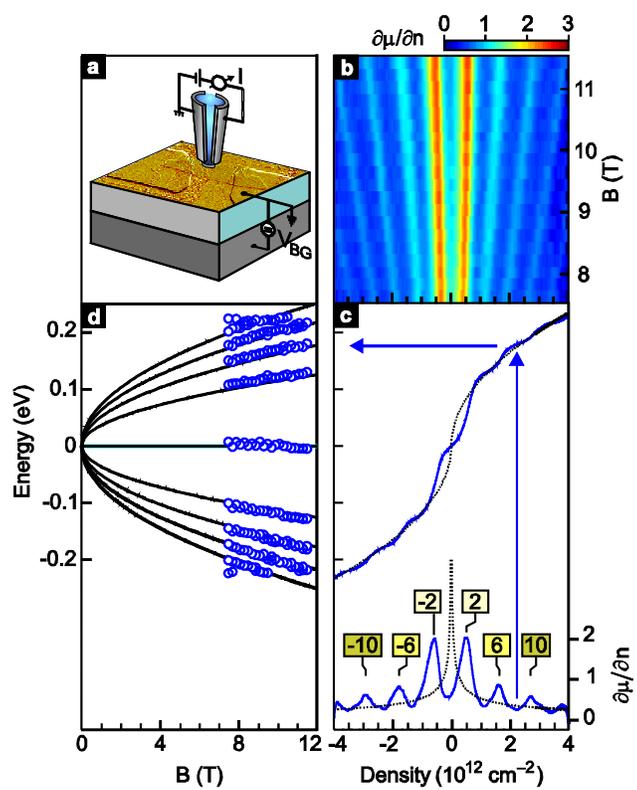

Figure 1

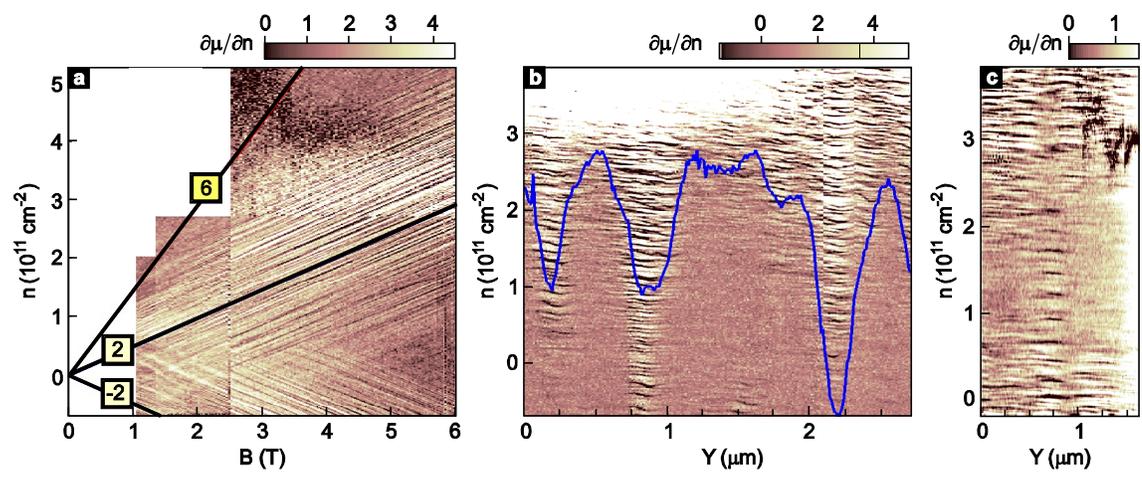

Figure 2

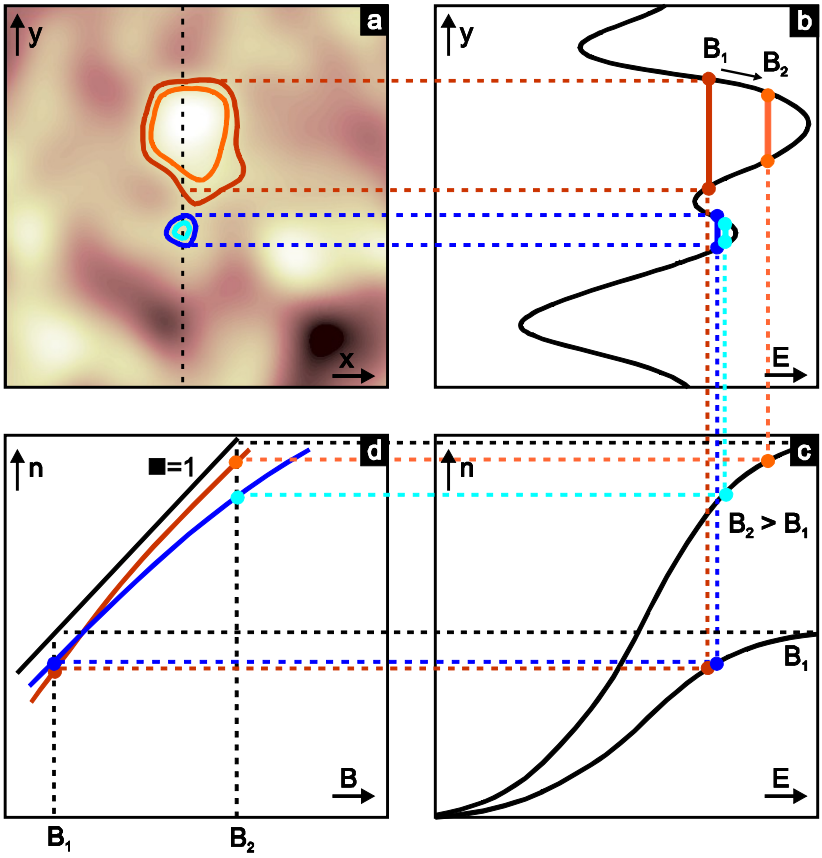

Figure 3

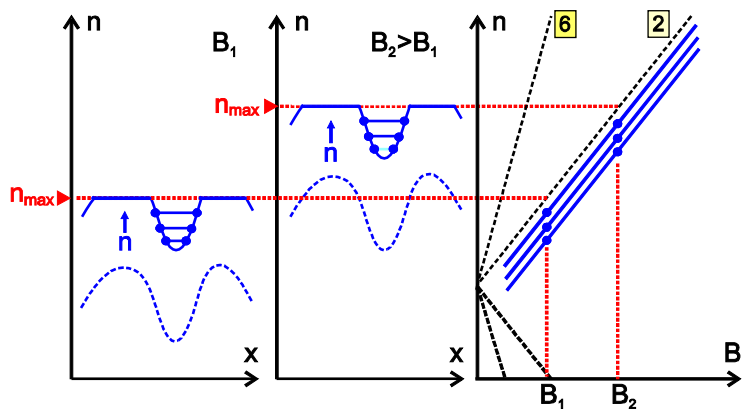

Figure 4

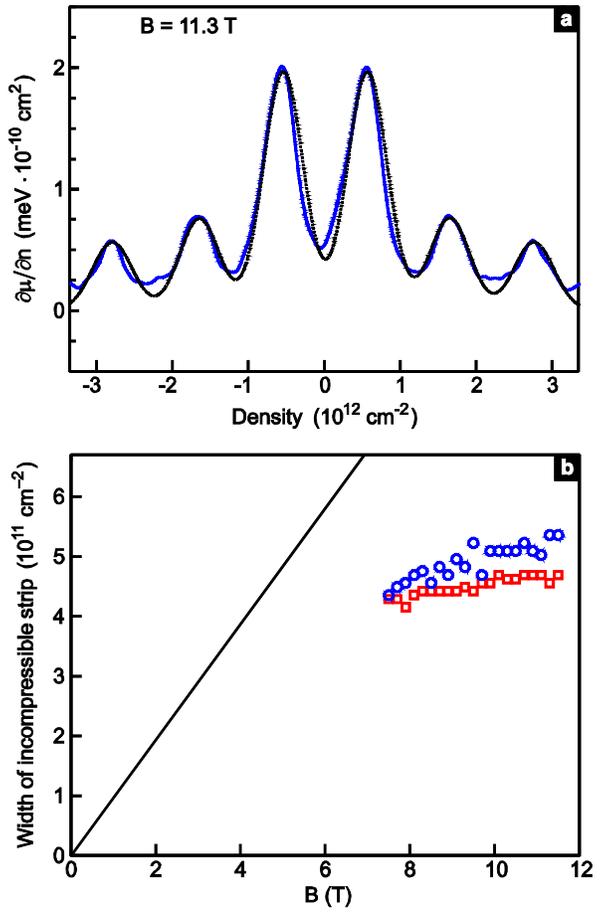

Figure 5